\documentclass{amsart} 
\usepackage{graphicx}
\usepackage{amssymb, amsmath, sidecap}
\usepackage{amsfonts}
\usepackage{amssymb}
\usepackage{float}
\usepackage[version=3]{mhchem}
\usepackage{endnotes}

\newcommand{\N}{\mathbb N}

\newcommand{\R}{\mathbb R}

\def\la{\label}

\def\bt{\begin{thm}}
\def\et{\end{thm}}
\def\bl{\begin{lem}}
\def\el{\end{lem}}
\def\bd{\begin{defi}}
\def\ed{\end{defi}}
\def\bc{\begin{cor}}
\def\ec{\end{cor}}
\def\bp{\begin{proof}}
\def\ep{\end{proof}}
\def\br{\begin{rem}}
\def\er{\end{rem}}

\def\bcon{\begin{conclusion}}
\def\econ{\end{conclusion}}

\newtheorem{thm}{Theorem}[section]
\newtheorem{lem}{Lemma}[section]
\newtheorem{defi}{Definition}[section]
\newtheorem{ex}{Example}[section]

\newtheorem{rem}{Remark}[section]
\newtheorem{cor}{Corollary}[section]

\newtheorem{conclusion}{Physical Conclusion}[section]

\numberwithin{equation}{section}
\numberwithin{figure}{section}

\begin{document}
\title{Phase Transitions for  the Brusselator Model}
\author[Ma]{Tian Ma}
\address[TM]{Department of Mathematics, Sichuan University,
Chengdu, P. R. China}

\author[Wang]{Shouhong Wang}
\address[SW]{Department of Mathematics,
Indiana University, Bloomington, IN 47405}
\email{showang@indiana.edu, http://www.indiana.edu/~fluid}

\thanks{The work was supported in part by grants from the
Office of Naval Research,  the National Science Foundation, and the National Science Foundation of China.}

\keywords{Belousov-Zhabotinsky chemical reactions, Brusselator, dynamic phase transition, spatiotemporal oscillations}
\subjclass{80A30, 35Q92}

\begin{abstract}
Dynamic phase transitions of the Brusselator model is carefully analyzed, leading to a rigorous characterization of the types and structure of the phase  transitions of the model from basic homogeneous states. 
The study is based on the dynamic transition theory developed recently by the authors.
\end{abstract}
\maketitle

\section{Introduction}
Belousov-Zhabotinsky (BZ) reactions are now one of a class of reactions that serve as a classical example of non-equilibrium thermodynamics, resulting in the establishment of a nonlinear chemical oscillator. 

The main objective of this article is to study the dynamic phase transitions of the Belousov-Zhabotinsky reactions focusing on   the Brusselator,
first introduced  by \cite{PL68}.  The Brusselator is one of the simplest models in nonlinear chemical systems. It has six components, four of which retain constants, and
the other two permit their concentrations vary with time and
space. The chemical reaction consists of four irreversible steps, given  by
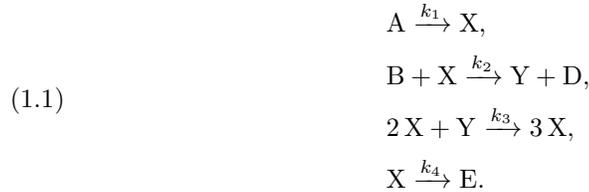
\begin{equation}
\begin{aligned}
& \ce{A ->[k_1]  X},\\
&\ce{B + X ->[k_2] Y + D},\\
&\ce{2X + Y ->[k_3] 3X},\\  
&\ce{X ->[k_4] E}.
\end{aligned}
\label{11.76}
\end{equation}
where $A$ and $B$ are constant components, $D$ and $E$  are  products, and
$X$  and $Y$ are the two components variable in time and space. Over the years, there have been extensive studies for the Brusselator and related chemical reaction problems; see among many others \cite{Matkowsky, guo-han, bao02, kaper,shi} and the references therein.

In this article, we address the dynamic phase transition of the Brusselator model. In particular, we derive a complete characterization of the transition from the homogeneous state. There are two aspects of this characterization.  First our analysis shows that both the transitions to multiple equilibria and to time-periodic solutions (spatiotemporal oscillations) can occur for the Brusselator model, and are precisely determined by the sign of an explicit  nondimensional parameter $\delta_0-\delta_1$ as defined by (\ref{11.88})  and (\ref{11.89}). 

Then in both transition cases, the dynamic behavior of the transition is classified based on the new dynamical classification scheme, introduced as part of the dynamical transition theory developed recently by the authors; see  \cite{ptd, b-book, chinese-book}. With this classification scheme, phase transitions are classified into three types: Type-I, Type-II and Type-III, which, in more mathematically intuitive terms,  are   called continuous, jump and mixed transitions respectively.   For the Brusselator, the distinction of the dynamic  transition types are determined, again,  by the signs of  some nondimensional computable parameters. 

It is worth mentioning that  the main philosophy of the dynamic transition theory is to search for  the full set of  transition states, giving a complete characterization on stability and  transition. The set of transition states is represented by a local attractor. Following this philosophy, the dynamic transition theory is developed  to identify the transition states and to classify them both dynamically and physically.  With this theory, many long standing phase transition problems are either solved or become more accessible, providing new insights to  both theoretical and experimental studies for the underlying physical problems.

This article is organized as follows. Section 2 introduces the model and its mathematical set-up, and Section 3 addresses the principle of  exchange of stabilities. Dynamic transitions of the model are addressed in Sections 4-6, with physical remarks of the main results given in Section 7.

\section{The Model and its Mathematical Set-up}

Let $u_1,u_2, a$ and $b$ stand for the concentrations of $X,Y,A$ and
$B$. Then the reaction equations of (\ref{11.76}) read
\begin{equation}
\begin{aligned}
&\frac{\partial u_1}{\partial t}=\sigma_1\Delta
u_1+k_1a-(k_2b+k_4)u_1+k_3u^2_1u_2,\\
&\frac{\partial u_2}{\partial t}=\sigma_2\Delta
u_2+k_2bu_1-k_3u^2_1u_2.
\end{aligned}
\label{11.77}
\end{equation}

To get the nondimensional form of (\ref{11.77}), let
\begin{align*}
&t=k^{-1}_4t^{\prime},  &&  x=lx^{\prime},  && 
   u_i=\left(\frac{k_4}{k_3}\right)^{{1}/{2}}u^{\prime}_i, \\
&a=(k^3_4/k^2_1k_3)^{{1}/{2}}\alpha , &&  b=(k_4/k_2)\lambda, && \sigma_i=l^2k_4\mu_i,
\end{align*}
for $i=1, 2$. Omitting the primes,  the equations (\ref{11.77}) become
\begin{equation}
\begin{aligned}
&\frac{\partial u_1}{\partial t}=\mu_1\Delta u_1+\alpha -(\lambda
+1)u_1+u^2_1u_2,\\
&\frac{\partial u_2}{\partial t}=\mu_2\Delta u_2+\lambda
u_1-u^2_1u_2,
\end{aligned}
\label{11.78}
\end{equation}
where $u_1,u_2\geq 0$,   $\Omega\subset \R^n$  $(1\leq n\leq 3)$ is a bounded
domain, and
$$\mu_1,\mu_2,\alpha ,\lambda >0.$$

The equations (\ref{11.78}) have a constant steady state solution
$$u_0=(\alpha ,\lambda /\alpha ).$$
Make the translation
$$u_1=\alpha +v_1,\ \ \ \ u_2=\frac{\lambda}{\alpha}+v_2,$$
then the equations (\ref{11.78}) are written as
\begin{equation}
\begin{aligned}
&\frac{\partial v_1}{\partial t}=\mu_1\Delta v_1+(\lambda
-1)v_1+\alpha^2v_2+\frac{2\lambda}{\alpha}v^2_1+2\alpha
v_1v_2+v^2_1v_2,\\
&\frac{\partial v_2}{\partial t}=\mu_2\Delta v_2-\lambda
v_1-\alpha^2v_2-\frac{2\lambda}{\alpha}v^2_1-2\alpha
v_1v_2-v^2_1v_2.
\end{aligned}
\label{11.79}
\end{equation}

There are two types of physically-sound boundary
conditions: the Dirichlet boundary condition
\begin{equation}
v=(v_1,v_2)=0\ \ \ \ \text{on}\ \partial\Omega ,\label{11.80}
\end{equation}
and the Neumann boundary condition
\begin{equation}
\frac{\partial v}{\partial n}=0\ \ \ \ \text{on}\ \partial\Omega
.\label{11.81}
\end{equation}

Define the  function spaces
\begin{eqnarray*}
&&H=L^2(\Omega ,\R^2),\\
&&H_1=\left\{
\begin{aligned} 
&H^2(\Omega ,\R^2)\cap H^1_0(\Omega,\R^2) &&  \text{for\ b.c}\ (\ref{11.80}),\\
&\{v\in H^2(\Omega ,\R^2)|\ \frac{\partial v}{\partial
n}|_{\partial\Omega}=0\} && \text{for\ b.c.}\ (\ref{11.81}).
\end{aligned}\right.
\end{eqnarray*}
Define the operators $L_{\lambda}=A+B_{\lambda}$ and
$G:H_1\rightarrow H$ by
\begin{equation}
\begin{aligned}
&Av=(\mu_1\Delta v_1,\mu_2\Delta v_2),\\
&B_{\lambda}v=((\lambda -1)v_1+\alpha^2v_2,-\lambda
v_1-\alpha^2v_2),\\
&Gv=\left(\frac{2\lambda}{\alpha}v^2_1+2\alpha
v_1v_2+v^2_1v_2,-\frac{2\lambda}{\alpha}v^2_1-2\alpha
v_1v_2-v^2_1v_2\right).
\end{aligned}
\label{11.82}
\end{equation}
Thus the equations (\ref{11.79}) with (\ref{11.80}) or with
(\ref{11.81}) can be written in the following abstract form
\begin{equation}
\frac{dv}{dt}=L_{\lambda}v+G(v, \lambda).\label{11.83}
\end{equation}

\section{Principle of Exchange of Stability (PES)}

Consider the eigenvalue quations of (\ref{11.79})
\begin{equation}
\begin{aligned}
&\mu_1\Delta v_1+(\lambda -1)v_1+\alpha^2v_2=\beta v_1,\\
&\mu_2\Delta v_2-\lambda v_1-\alpha^2v_2=\beta v_2,
\end{aligned}
\label{11.84}
\end{equation}
with the boundary condition (\ref{11.80}) or (\ref{11.81}).

Let $\rho_k$ and $e_k$ be the $k$th eigenvalue and eigenvector of the Laplacian with either the Dirichlet or the Neumann condition:
 \begin{equation}
\begin{aligned} 
&\Delta e_k=-\rho_ke_k,\\
&e_k|_{\partial\Omega}=0 \qquad \text{or}\ \frac{\partial e_k}{\partial
n}|_{\partial\Omega}=0.
\end{aligned}
\label{11.85}
\end{equation}
Denote by $M_k$ the matrix given by
$$M_k=\left(\begin{array}{cc}
-\mu_1\rho_k+\lambda -1&\alpha^2\\
-\lambda&-\mu_2\rho_k-\alpha^2
\end{array}
\right),\ \ \ \ k=1,2,\cdots .$$

It is clear that all eigenvalues $\beta^{\pm}_k$ and eigenvectors
$\phi^{\pm}_k$ of (\ref{11.84}) satisfy the following equations
\begin{equation}
\begin{aligned}
&\phi^{\pm}_k=\xi^{\pm}_ke_k,\\
&M_k\xi^{\pm}_k=\beta^{\pm}_k\xi^{\pm}_k,
\end{aligned}
\label{11.86}
\end{equation}
where $\xi^{\pm}_k\in \R^2$ are the eigenvectors of $M_k,
\beta^{\pm}_k$ are the eigenvalues of $M_k$, which are expressed as
\begin{align}
\beta^{\pm}_k(\lambda )= & \frac{1}{2}[\lambda
-(\mu_1\rho_k+\mu_2\rho_k+\alpha^2+1)]\label{11.87}\\
&\pm\frac{1}{2}\big[(\lambda
-\mu_1\rho_k-\mu_2\rho_k-\alpha^2-1)^2 \nonumber  \\
& +4(\lambda\mu_2\rho_k-(\mu_1\rho_k+1)(\mu_2\rho_k+
\alpha^2))\big]^{{1}/{2}}.\nonumber 
\end{align} 
It is clear  that $\beta^-_k(\lambda )<\beta^+_k(\lambda
)=0$ if and only if
\begin{eqnarray*}
&&\lambda
=\frac{1}{\mu_2\rho_k}(\mu_1\rho_k+1)(\mu_2\rho_k+\alpha^2),\\
&&\lambda <\mu_1\rho_k+\mu_2\rho_k+\alpha^2+1,
\end{eqnarray*}
and $\beta^{\pm}_k(\lambda )=\pm\sigma_k(\lambda )i$ with
$\sigma_k\neq 0$ if and only if
\begin{eqnarray*}
&&\lambda =\mu_1\rho_k+\mu_2\rho_k+\alpha^2+1,\\
&&\lambda
<\frac{1}{\mu_2\rho_k}(\mu_1\rho_k+1)(\mu_2\rho_k+\alpha^2).
\end{eqnarray*}

Thus we introduce two critical numbers
\begin{align}
& \lambda_0=\min_{\rho_k}\frac{1}{\mu_2\rho_k}(\mu_1\rho_k+1)(\mu_2\rho_k+\alpha^2),
\label{11.88}
\\\
&
\lambda_1=\mu_1\rho_1+\mu_2\rho_1+\alpha^2+1.\label{11.89}
\end{align}
Obviously, the following lemma holds true.

\bl\la{l11.3}
 Let $\lambda_0$ and $\lambda_1$ be the two
numbers given by (\ref{11.88}) and (\ref{11.89}). Then we have the
following assertions:

\begin{itemize}

\item[(1)] Let $\lambda_0<\lambda_1$, and $k_0\geq 1$ be the
integer such that the minimum is achieved at $\rho_{k_0}$ in the definition of  $\lambda_0$ . Then
$\beta^+_{k_0}(\lambda )$ is the first real eigenvalue of
(\ref{11.84}) near $\lambda =\lambda_0$ satisfying that
\begin{equation}
\begin{aligned}
&\beta^+_k(\lambda )
\left\{\begin{array}{ll}
<0 & \text{ if } \lambda <\lambda_0\\
=0 &\text{ if } \lambda =\lambda_0\\
>0 &\text{ if } \lambda >\lambda_0
\end{array}\right. &&  \forall  k \in \N \text{  with}\ \rho_k=\rho_{k_0},\\
&\text{Re}\beta^{\pm}_j(\lambda_0)<0 &&
\forall\beta^{\pm}_j\neq\beta^+_k\ \text{with}\ \rho_k=\rho_{k_0}.
\end{aligned}
\label{11.90}
\end{equation}
\item[(2)] Let $\lambda_1<\lambda_0$. Then $\beta^+_1(\lambda
)=\bar{\beta}^-_1(\lambda )$ are a pair of first complex eigenvalues
of (\ref{11.84}) near $\lambda =\lambda_1$, and
\begin{equation}
\begin{aligned}
&\text{Re}\beta^+_1(\lambda )=\text{Re}\beta^-_1(\lambda
)\left\{\begin{array}{ll} <0&  \text{ if } \lambda <\lambda_1,\\
=0& \text{ if } \lambda =\lambda_1,\\
>0& \text{ if } \lambda >\lambda_1,
\end{array}
\right.\\
&\text{Re}\beta^{\pm}_k(\lambda_1)<0 \qquad \quad \forall k>1.
\end{aligned}
\label{11.91}
\end{equation}
\end{itemize}
\el

\br\la{r11.4}
{\rm
 $\beta^{\pm}_1(\lambda )$ are simple complex
eigenvalues at $\lambda_1(<\lambda_0)$, and in general, if
$\rho_{k_0}$ is a simple eigenvalue of (\ref{11.85}), then
$\beta^+_{k_0}(\lambda )$ is also simple at $\lambda_0(<\lambda_1)$.
}
\er

\section{Transition from real eigenvalues}

Hereafter, we always assume that the eigenvalue $\beta^+_{k_0}$ in
(\ref{11.90}) is simple. Based on Lemma \ref{l11.3}, as
$\lambda_0<\lambda_1$ the transition of (\ref{11.83}) occurs at
$\lambda =\lambda_0$, which is from real eigenvalues. Let
$\rho_{k_0}$ be as in Lemma \ref{l11.3}, and $e_{k_0}$ the eigenvector of
(\ref{11.85}) corresponding to $e_{k_0}$ satisfying
\begin{equation}
\int_{\Omega}e^3_{k_0}dx\neq 0.\label{11.92}
\end{equation}

Then, under the condition (\ref{11.92}), for the system
(\ref{11.79}) with (\ref{11.80}) or with (\ref{11.81}) we have the
following transition theorem.

\bt\la{t11.6}
 Let $\lambda_0<\lambda_1$. Then the system
(\ref{11.83}) has a transition at $\lambda =\lambda_0$, which is
mixed (Type-III). In particular, the system bifurcates on each side
of $\lambda =\lambda_0$ to a unique branch $v^{\lambda}$ of steady
state solutions, such that the following assertions hold true:

\begin{itemize}
\item[(1)]  On $\lambda <\lambda_0$, the bifurcated solution
$v^{\lambda}$ is a saddle, and the stable manifold
$\Gamma^1_{\lambda}$ of $v^{\lambda}$ separates the space $H$ into
two disjoint open sets $U^{\lambda}_1$ and $U^{\lambda}_2$, such
that $v=0\in U^{\lambda}_1$ is an attractor, and the orbits of
(\ref{11.83}) in $U^{\lambda}_2$ are far from $v=0$.

\item[(2)] On $\lambda >\lambda_0$, the stable manifold
$\Gamma^0_{\lambda}$ of $v=0$ separates the neighborhood $\mathcal O$ of
$u=0$ into two disjoint open sets $\mathcal O^{\lambda}_1$ and
$\mathcal O^{\lambda}_2$, such that the transition is
jump in $\mathcal O^{\lambda}_1$,  and is continuous  in $O^{\lambda}_2$. The bifurcated
solution $v^{\lambda}\in \mathcal O^{\lambda}_2$ is an attractor such that
for any $\varphi\in \mathcal O^{\lambda}_2,$
$$\lim\limits_{t\rightarrow\infty}\|v(t,\varphi
)-v^{\lambda}\|_H=0,$$ where $v(t,\varphi )$ is the solution of
(\ref{11.83}) with $v(0,\varphi )=\varphi $.

\item[(3)] The bifurcated solution $v^{\lambda}$ can be expressed
as
\begin{equation}
\begin{aligned}
&v^{\lambda}=C\beta^+_{k_0}(\lambda
)\xi^+_{k_0}e_{k_0}+o(|\beta^{\pm}_{k_0}|),\\
&\xi^+_{k_0}=(-\mu_2\rho_{k_0},\mu_1\rho_{k_0}+1),\\
&C=\frac{(\alpha
\mu_2\rho_{k_0}(\mu_2\rho_{k_0}+\alpha^2)-\alpha^3(\mu_1\rho_{k_0}+1))\int_{\Omega}e^2_{k_0}dx}
{2\mu^3_2\rho^3_{k_0}(\mu_1\rho_{k_0}+1)\int_{\Omega}e^3_{k_0}dx}.
\end{aligned}
\label{11.93}
\end{equation}
\end{itemize}
\et

\bp 
We apply Theorem~A.2 in \cite{MW08a} 
 to prove this theorem. Let
$\Phi$ be the center manifold function of (\ref{11.83}) at $\lambda
=\lambda_0$. We need to simplify the following expression:
\begin{equation}
g(y)=\frac{1}{(\phi^+_{k_0},\phi^{+*}_{k_0} )}(G(y\phi^+_{k_0}+\Phi
(y)),\phi^{+*}_{k_0} ),\label{11.94}
\end{equation}
where $y\in \R^1$, $G$ is the operator defined by (\ref{11.82}),
$\phi^+_{k_0}$  is the eigenvector of (\ref{11.81}) corresponding to
$\beta^+_{k_0}(\lambda_0)=0$, and $\phi^{+*}_{k_0}$  is  the conjugate
eigenvector. By (\ref{11.86}),  
\begin{align}
&  \phi^+_{k_0}=\xi^+_{k_0}e_{k_0},\ \ \ \
\phi^{+*}_{k_0}=\xi^{+*}_{k_0}e_{k_0},\label{11.95}
\\
&\left(\begin{array}{cc} \lambda_0-(\mu_1\rho_{k_0}+1)&\alpha^2\\
-\lambda_0&-(\mu_2\rho_{k_0}+\alpha^2)
\end{array}\right)\left(\begin{array}{l}
\xi^+_{k_0l}\\
\xi^+_{k_02}
\end{array}\right)=0,\label{11.96}\\
&\left(\begin{array}{cc} \lambda_0-(\mu_1\rho_{k_0}+1)&-\lambda_0\\
\alpha^2&-(\mu_2\rho_{k_0}+\alpha^2)
\end{array}\right)\left(\begin{array}{l}
\xi^{+*}_{k_0l}\\
\xi^{+ *}_{k_02}
\end{array}\right)=0.
\label{11.97}
\end{align}
By definition of 
$\lambda_0$  and $k_0$,  we infer from (\ref{11.96}) and (\ref{11.97}) that 
\begin{eqnarray}
&&\xi^+_{k_0}=(\xi^+_{k_01},\xi^+_{k_02})=(-\mu_2\rho_{k_0},\mu_1\rho_{k_0}+1),\label{11.98}\\
&&\xi^{+*}_{k_0}=(\xi^{+*}_{k_01},\xi^{+*}_{k_02})=(\mu_2\rho_{k_0}+\alpha^2,\alpha^2).\label{11.99}
\end{eqnarray}
Denote by $o(k)=o(|y|^k)$. By $\Phi (y)=o(1)$, the function $g(y)$
in (\ref{11.94}) is rewritten as
$$g(y)=\frac{1}{(\phi^+_{k_0},\phi^{+*}_{k_0} )}(G(y\phi^+_{k_0}),\phi^{+*}_{k_0})+o(2).$$
By (\ref{11.95}) and (\ref{11.98}) we see that
$$G(y\phi^+_{k_0})=\left\{\begin{aligned}
&2y^2\left(\frac{\lambda_0}{\alpha}\mu^2_2\rho^2_k-\alpha\mu_2\rho_{k_0}(\mu_1\rho_{k_0}+1)\right)
e^2_+{k_0}+o(2),\\
&-2y^2\left(\frac{\lambda_0}{\alpha}\mu^2_2\rho^2_k-\alpha\mu_2\rho_{k_0}(\mu_1\rho_{k_0}+1)
\right)e^2_{k_0}+o(2).
\end{aligned}
\right.
$$ 
Thus   we deduce from (\ref{11.95}) and (\ref{11.98})-(\ref{11.99})
that
\begin{eqnarray*}
&&(\phi^+_{k_0},\phi^{+*}_{k_0} )=(\alpha^2(\mu_1\rho_{k_0}+1)-\mu_2\rho_{k_0}(\mu_2\rho_{k_0}
+\alpha^2)\int_{\Omega}e^2_{k_0}dx\\
&&(G(y\phi^+_{k_0},\phi^{+*}_{k_0} ) =y^2\cdot\frac{2\mu^3_2\rho^3_{k_0}}{\alpha}(\mu_1
\rho_{k_0}+1)\int_{\Omega}e^3_{k_0}dx+o(2).
\end{eqnarray*}
Therefore the function (\ref{11.94})  is given by as
$$g(y)=-\frac{1}{c}y^2+o(2),$$
and the theorem follows from Theorem~A.2 in \cite{MW08a}. 
The proof is
complete.\ep

\br\la{r11.5}
{\rm
 If the domain $\Omega\neq (0,L)\times D$ with
$D\subset \R^2$ being a bounded open set, then the condition (\ref{11.92})
holds true for both the Dirichlet and Neumann boundary conditions (\ref{11.80}) and
(\ref{11.81}). If $\Omega =(0,L)\times D$, then (\ref{11.92}) is
not true for   the Neumann condition (\ref{11.81}), and is not true  for the Dirichlet condition (\ref{11.80})
if  the number $m$ in
$\rho_{k_0}=\frac{m^2\pi^2}{L^2}+\rho^{\prime}_{k_0}$ is  even. Here 
$\rho^{\prime}_{k_0}$ is an eigenvalue of the equation on $D$:
$$\begin{array}{l}
-\Delta e=\rho^{\prime}_{k_0}e\qquad \text{ for }  x\in D,\\
e|_{\partial D}=0.
\end{array}
$$
\qed }
\er

Now, we consider the case where (\ref{11.92}) is not true, i.e.,
\begin{equation}
\int_{\Omega}e^3_{k_0}dx=0.\label{11.100} \end{equation} As
mentioned in Remark \ref{r11.5}, the condition (\ref{11.100}) may hold  true if
$\Omega =(0,L)\times D$. We introduce the following parameter
\begin{align}
b_1=  & \big[
  \alpha\mu^2_2\rho^2_{k_0}(\mu_1\rho_{k_0}+1)\int_{\Omega}e^4_{k_0}dx-2\alpha^2\mu_2
\rho_{k_0}\int_{\Omega}\psi_2e^2_{k_0}dx\label{11.101}\\
& -2(\mu_1\rho_{k_0}+1)(2\mu_2\rho_{k_0}+\alpha^2)\int_{\Omega}\psi_1e^2_{k_0}dx\big] \nonumber\\
& \times
[\alpha^2(\mu_1\rho_{k_0}+1)-\mu_2\rho_{k_0}(\mu_2\rho_{k_0}+\alpha^2)]^{-1}, \nonumber 
\end{align}
where $\psi =(\psi_1,\psi_2)$ satisfies
\begin{equation}
\begin{aligned}
&\mu_1\Delta\psi_1+(\lambda_0-1)\psi_1+\alpha^2\psi_2=-\frac{2\mu^2_2\rho^2_{k_0}}{\alpha}
(\mu_1\rho_{k_0}+1)e^2_{k_0},\\
&\mu_2\Delta\psi_2-\lambda_0\psi_1-\alpha^2\psi_2=\frac{2\mu^2_2\rho^2_{k_0}}{\alpha}(\mu_1
\rho_{k_0}+1)e^2_{k_0},\\
&\psi |_{\partial\Omega}=0\ \ \ \ (\text{or}\
\frac{\partial\psi}{\partial n}|_{\partial\Omega}=0).
\end{aligned}\label{11.102}
\end{equation}
By the Fredholm Alternative Theorem, under the condition
(\ref{11.100}), the equation (\ref{11.102}) has a unique solution.

\bt\la{t11.7}
 Let (\ref{11.100}) hold true,
$\lambda_0<\lambda_1$, and $b_1$ is the number given by (\ref{11.101}).
Then the transition of (\ref{11.83}) at $\lambda =\lambda_0$ is
continuous if  $b_1<0$, and is jump if  $b_1>0$. Moreover, the following assertions hold true:

\begin{itemize}
\item[(1)]  If $b_1>0$, (\ref{11.83}) has no bifurcation on $\lambda
>\lambda_0$, and has exact two bifurcated solutions $v^{\lambda}_+$
and $v^{\lambda}_-$ which are saddles. Moreover,  the stable manifolds
$\Gamma^{\lambda}_+$ and $\Gamma^{\lambda}_-$ of the two bifurcated solutions divide $H$ into three
disjoint open sets $U^{\lambda}_+$, $U^{\lambda}_0$, $U^{\lambda}_-$ such
that $v=0\in U^{\lambda}_0$ is an attractor, and the orbits in
$U^{\lambda}_{\pm}$ are far from $v=0$.

\item[(2)]   If $b_1<0$, (\ref{11.83}) has no bifurcation on $\lambda
<\lambda_0$, and has exact two bifurcated solutions $v^{\lambda}_+$
and $v^{\lambda}_-$, which are attractors. In addition, there is a
neighborhood $\mathcal O\subset H$ of $v=0$, such that the stable manifold $\Gamma$ of
$v=0$ divides $\mathcal O$ into two disjoint open sets $\mathcal O^{\lambda}_+$ and
$\mathcal O^{\lambda}_-$ such that $v^{\lambda}_+\in
\mathcal O^{\lambda}_+$,  $v^{\lambda}_-\in \mathcal O^{\lambda}_-$, and
$v^{\lambda}_{\pm}$ attracts $\mathcal O^{\lambda}_{\pm}$;

\item[(3)]  The bifurcated solutions $v^{\lambda}_{\pm}$ can be
expressed as
\begin{equation}
\begin{aligned}
&v^{\lambda}_{\pm}=\pm C(\beta^{+}_{k_0}(\lambda))^{1/2} 
 \xi^+_{k_0}e_{k_0}+o(|\beta^+_{k_0}|^{{1}/{2}}),\\
&\xi^+_{k_0}=(-\mu_2\rho_{k_0},\mu_1\rho_{k_0}+1),\\
&C=\left[\frac{-\alpha}{\mu_2\rho_{k_0}b_1}\int_{\Omega}e^2_{k_0}dx\right]^{{1}/{2}},
\end{aligned}
\label{11.103}
\end{equation}
where $b_1$ is as in (\ref{11.101}).
\end{itemize}
\et

\bp
 We use Theorem~A.1 in \cite{MW08a} 
 to prove this theorem. To get the
function $g(y)$ in (\ref{11.94}), we need to calculate the center
manifold function $\Phi (y)$. By (A.10) in \cite{MW09c}, 
$\Phi (y)$ satisfies
\begin{equation}
L_{\lambda_0}\Phi =-P_2G(y\phi^+_{k_0}),\label{11.104}
\end{equation}
where $P_2:H\rightarrow E_2$ is the canonical projection,
$L_{\lambda}$  is  as in (\ref{11.82}), $\phi^+_{k_0}$ and $\phi^{+*}_{k_0}$ are given by (\ref{11.95}), 
and
$$
E_2=\{v\in H|\ (v,\phi^{+*}_{k_0} )=0\}.
$$
We  see that
$$G(y\phi^+_{k_0})=\left\{\begin{aligned}
&2\alpha^{-1}\mu^2_2\rho^2_{k_0}(\mu_1\rho_{k_0}+1)y^2e^2_{k_0}+o(2), \\
&-2\alpha^{-1}\mu^2_2\rho^2_{k_0}(\mu_1\rho_{k_0}+1)y^2e^2_{k_0}+o(2).
\end{aligned}
\right.$$ 
Let 
\begin{equation}
\Phi =\psi y^2+o(2),\ \ \ \ \psi =(\psi_1,\psi_2)\in
H.\label{11.105}
\end{equation}
By (\ref{11.100}), $(e^2_{k_0},-e^2_{k_0})\in E_2$. Hence, it
follows from (\ref{11.104}) and (\ref{11.105}) that 
\begin{equation}
L_{\lambda_0}\psi
=-2\alpha^{-1}\mu^2_2\rho^2_{k_0}(\mu_1\rho_{k_0}+1)(e^2_{k_0},-e^2_{k_0}),\label{11.106}
\end{equation}
which is an equivalent form of (\ref{11.102}).

By (\ref{11.105}), we have
\begin{align*}
&G(y\phi^+_{k_0}+\Phi )=G(y\phi^+_{k_0}+y^2\psi )+o(3),\\
&G(y\phi^+_{k_0}+y^2\psi )
=\left\{\begin{aligned}
   & O(y^2)e^2_{k_0}+y^3[\mu^2_2\rho^2_{k_0})(\mu_1\rho_{k_0}+1)e^3_{k_0}-2\alpha\mu_2\rho_{k_0}\psi_2
e_{k_0}\\
   & \qquad -2\alpha^{-1}\psi_1e_{k_0}(\mu_1\rho_{k_0}+1)(2\mu_2\rho_{k_0}+\alpha^2)]+o(3),\\
   & O(y^2)e^2_{k_0}-y^3[\mu^2_2\rho^2_{k_0}(\mu_1\rho_{k_0}+1)e^3_{k_0}-2\alpha\mu_2\rho_{k_0}\psi_2e_{k_0}\\
   & \qquad -2\alpha^{-1}\psi_1e_{k_0}(\mu_1\rho_{k_0}+1)(2\mu_2\rho_{k_0}+\alpha^2)]+o(3).
\end{aligned} \right.
\end{align*}
Hence we deduce  from (\ref{11.95}), (\ref{11.99}) and (\ref{11.100}) that
\begin{align*}
(G(y\phi^+_{k_0}+\Phi),\Phi^{+*}_{k_0} ) =
  &  \frac{\mu_2\rho_{k_0}y^3}{\alpha}\Big[\alpha\mu^2_2\rho^2_{k_0}(\mu_1\rho_{k_0}+1)
\int_{\Omega}e^4_{k_0}dx -2\alpha^2\mu_2\rho_{k_0}\int_{\Omega}\psi_2e^2_{k_0}dx \\ 
&  -2(\mu_1\rho_{k_0}+1)(2\mu_2\rho_{k_0}+
\alpha^2)\int_{\Omega}\psi_1e^2_{k_0}dx\Big]+o(3). 
\end{align*} 
Thus,
the function $g(y)$ in (\ref{11.94}) can be written as
$$g(y)=\frac{\mu_2\rho_{k_0}b_1}{\alpha\int_{\Omega}e^2_{k_0}dx}y^3+o(3),$$
where $b_1$ is as in (\ref{11.101}). Hence the theorem follows from
Theorem~A.1 in \cite{MW08a}. 
\ep

When the domain $\Omega$ is a rectangle, i.e. $\Omega
=\prod\limits^n_{j=1}(0,L_j)$, the $b_1$  in (\ref{11.101}) for the Neumann condition can be explicitly expressed in terms of  the
physical parameters $\mu_1,\mu_2,\alpha$, and $L_j$   $(1\leq j\leq n)$.
For example, we consider the case where $\Omega =(0,L)$. The eigenvalues $\rho_k$ and eigenvectors $e_k$ of (\ref{11.85})  are given by
$$\rho_k=\frac{(k-1)^2\pi^2}{L^2},\ \ \ \
e_k=\cos\frac{(k-1)\pi}{L}x,\ \ \ \ k=1,2,\cdots .$$ It is clear
that $k_0\geq 2$, and (\ref{11.100}) holds true. We see that
$$
e^2_{k_0}= \frac{1}{2}(1+\cos\frac{2(k_0-1)\pi}{L}x)= \frac{1}{2}[e_1+ e_{2k_0-1}].
$$
Hence, by (\ref{11.102}),  we have
\begin{equation}
\psi =\xi e_1+\eta e_j \qquad \text{ with } j=2k_0-1,\label{11.107}
\end{equation}
where 
\begin{align*}
&
\left(\begin{matrix}
 \lambda_0-1&\alpha^2\\ 
 -\lambda_0&-\alpha^2
\end{matrix} \right)
    \left(\begin{matrix} \xi_1\\
         \xi_2
          \end{matrix} \right)
  =\left(\begin{matrix}
   -\frac{1}{\alpha}\mu^2_2\rho^2_{k_0}(\mu_1\rho_{k_0}+1)\\
    \frac{1}{\alpha}\mu^2_2\rho^2_{k_0}(\mu_1\rho_{k_0}+1)
\end{matrix}\right), 
  \\
&
\left(\begin{matrix} 
 \lambda_0-(\mu_1\rho_j+1)&\alpha^2\\
-\lambda_0&-(\mu_2\rho_j+\alpha^2)
\end{matrix}\right)
\left(\begin{matrix}
\eta_1\\
\eta_2
\end{matrix}\right)
=\left(\begin{matrix}
-\frac{1}{\alpha}\mu^2_2\rho^2_{k_0}(\mu_1\rho_{k_0}+1)\\
\frac{1}{\alpha}\mu^2_2\rho^2_{k_0}(\mu_1\rho_{k_0}+1)
\end{matrix}\right), 
\\
&
\lambda_0=\frac{1}{\mu_2\rho_{k_0}}(\mu_1\rho_{k_0}+1)(\mu_2\rho_{k_0}+\alpha^2).
\end{align*}
It is readily to see that 
\begin{equation}
\begin{aligned}
&\xi_1=0,\\
&\xi_2=-\alpha^{-3}\mu^2_2\rho^2_{k_0}(\mu_1\rho_{k_0}+1),\\
&\eta_1=\frac{\mu^3_2\rho_j\rho^2_{k_0}(\mu_1\rho_{k_0}+1)}{\alpha
[(\mu_2\rho_j+\alpha^2)(\mu_1\rho_j+1)-\mu_2\rho_j\lambda_0]},\\
&\eta_2=\frac{-\mu^2_2\rho^2_{k_0}(\mu_1\rho_j+1)(\mu_1\rho_{k_0}+1)}{\alpha
[(\mu_2\rho_j+\alpha^2)(\mu_1\rho_j+1)-\mu_2\rho_j\lambda_0]}.
\end{aligned}
\label{11.108}
\end{equation}
Inserting (\ref{11.107}) and (\ref{11.108}) into (\ref{11.101}),  we
derive that
\begin{align}
b_1= &  
 \frac{\mu^2_2\rho^2_{k_0}(\mu_1\rho_{k_0}+1)L}{\alpha}\Big[\mu_2\rho_{k_0}+\frac{3}{8}\alpha^2
   \label{11.109}\\
&
+ \frac{\alpha^2(\mu_1\rho_j+1)\mu_2\rho_{k_0}-\frac{1}{2}(\mu_1\rho_{k_0}+1)(2\mu_2
\rho_{k_0}+\alpha^2)\mu_2\rho_j}{(\mu_1\rho_j+1)(\mu_2\rho_j+\alpha^2)-\mu_2\rho_j\lambda_0}\Big]\nonumber\\
& 
\times\left[\alpha^2(\mu_1\rho_{k_0}+1)-\mu_2\rho_{k_0}(\mu_2\rho_{k_0}+\alpha^2)\right]^{-1},\nonumber\\
\rho_{k_0}= & \frac{(k_0-1)^2\pi^2}{L^2},\nonumber\\
\rho_j=& \frac{4(k_0-1)^2\pi^2}{L^2}=4\rho_{k_0}.\nonumber
\end{align}
Thus, for the one dimensional domain $\Omega =(0,L)$, the number
$b_1$ in (\ref{11.109}) can be equivalently rewritten as
\begin{eqnarray}
&&b_1=\left[\mu_2\rho_{k_0}+\frac{3}{8}\alpha^2-\frac{\mu_2\rho_{k_0}(4\mu_1\mu_2\rho^2_{k_0}+
4\mu_2\rho_{k_0}-2\alpha^2\mu_1\rho_{k_0}+\alpha^2)}{12\mu_1\mu_2\rho^2_{k_0}-3\alpha^2}
\right]\label{11.110}\\
&&\ \
\times\left[\alpha^2(\mu_1\rho_{k_0}+1)-\mu_2\rho_{k_0}(\mu_2\rho_{k_0}+\alpha^2)\right]^{-1}.\nonumber
\end{eqnarray}

\section{Transition from complex eigenvalues}

As $\lambda_1<\lambda_0$, the transition of (\ref{11.83}) occurs at
$\lambda =\lambda_1$, and the system bifurcates to a periodic
solution.

We first consider the Neumann boundary condition. In this case,
$\lambda_1=\alpha^2+1$, and we have the following theorem.

\bt\la{t11.8}
 For the problem (\ref{11.79}) with
(\ref{11.81}), when $\lambda_1<\lambda_0$, the transition at
$\lambda =\lambda_1$ is continuous, and the problem bifurcates on
$\lambda >\lambda_1$ to one periodic solution which is an attractor.
Moreover, the bifurcated periodic solution
$v^{\lambda}=(v^{\lambda}_1,v^{\lambda}_2)$ can be expressed as
\begin{equation}
\begin{aligned}
&v^{\lambda}_1=2[C\beta_0(\lambda )]^{{1}/{2}}\alpha^2\sin
(\alpha t+\frac{\pi}{4})+o(|\beta_0|^{{1}/{2}}),\\
&v^{\lambda}_2=[C\beta_0(\lambda )]^{{1}/{2}}\alpha ((\alpha
-1)\cos\alpha t-(\alpha +1)\sin\alpha t)+o(|\beta_0|^{{1}/{2}}),
\end{aligned}
\label{11.111}
\end{equation}
where $\beta_0=\frac{1}{2}(\lambda -\lambda_1)$, and
$C=(2\pi\alpha^2+\frac{3}{2}\pi\alpha^4)^{-1}$.
\et

\bp
 We shall verify this theorem by using Theorem~A.3 in \cite{MW10a}.
The eigenvalue $\beta^{\pm}_1(\lambda )$ in (\ref{11.87})  are given by
$$\beta^{\pm}_1(\lambda )=\frac{1}{2}(\lambda
-\lambda_1)\pm\frac{i}{2}\sqrt{4\alpha^2-(\lambda -\lambda_1)^2},\ \
\ \ (\lambda_1=\alpha^2+1).
$$ 
Namely,  for   $ \lambda$   near $ \lambda_1$, 
\begin{align*}
&
\beta_0(\lambda )=\text{Re}\beta^{\pm}_1(\lambda)=\frac{1}{2}(\lambda -\lambda_1), \\
&
\beta^{\pm}_1(\lambda_1)=\pm i\alpha .
\end{align*}
The eigenvectors $\xi$ and $\eta$ corresponding to
$\beta^{\pm}_1(\lambda_1)$ satisfy
\begin{eqnarray*}
&&\left(\begin{array}{cc} \alpha^2&\alpha^2\\
-(\alpha^2+1)&-\alpha^2
\end{array}\right)\left(\begin{array}{l}
\xi_1\\
\xi_2
\end{array}\right)=\alpha\left(\begin{array}{c}
\eta_1\\
\eta_2
\end{array}\right),\\
&&\left(\begin{array}{cc} \alpha^2&\alpha^2\\
-(\alpha^2+1)&-\alpha^2
\end{array}\right)\left(\begin{array}{l}
\eta_1\\
\eta_2
\end{array}\right)=-\alpha\left(\begin{array}{l}
\xi_1\\
\xi_2
\end{array}\right),
\end{eqnarray*}
It is easy to see that \begin{equation}
\begin{aligned}
&\xi =(\xi_1,\xi_2)=(\alpha^2,\alpha (1-\alpha )),\\
&\eta =(\eta_1,\eta_2)=(\alpha^2,-\alpha (\alpha +1)).
\end{aligned}
\label{11.112}
\end{equation}
The conjugate eigenvectors $\xi^*$ and $\eta^*$ satisfy
\begin{eqnarray*}
&&\left(\begin{array}{cc} \alpha^2&-(\alpha^2+1)\\
\alpha^2&-\alpha^2 \end{array} \right)\left(\begin{array}{l}
\xi^*_1\\
\xi^*_2
\end{array}\right)=\alpha\left(\begin{array}{l}
\eta^*_1\\
\eta^*_2
\end{array}\right),\\
&&\left(\begin{array}{cc} \alpha^2&-(\alpha^2+1)\\
\alpha^2&-\alpha^2
\end{array}
\right)\left(\begin{array}{l} \eta^*_1\\
\eta^*_2
\end{array}\right)=-\alpha\left(\begin{array}{l}
\xi^*_1\\
\xi^*_2
\end{array}
\right),
\end{eqnarray*}
leading to
\begin{equation}
\begin{aligned}
&\xi^*=(\xi^*_1,\xi^*_2)=(\alpha (\alpha +1),\alpha^2),\\
&\eta^*=(eta^*_1,\eta^*_2)=(\alpha (\alpha -1),\alpha^2).
\end{aligned}
\label{11.113}
\end{equation}
It is readily to check that
\begin{equation}
\begin{aligned}
&(\xi ,\xi^* )=-(\eta,\eta^* ) =2\alpha^3\int_{\Omega}e^2_1dx=2\alpha^3|\Omega |,\\
&(\xi ,\eta^*)=(\eta ,\xi^* ) =0.
\end{aligned}
\label{11.114}
\end{equation}
For the operator $G$ defined by (\ref{11.82}),  we deduce  from (\ref{11.112})
that for $x,y\in \R^1$, 
\begin{align}
G(x\xi  &  +y\eta ) \label{11.115}\\
& = \left\{\begin{aligned}
                 &
 \frac{2\lambda_1}{\alpha}(x\xi_1+y\eta_1)^2+2\alpha
(x\xi_1+y\eta_1)(x\xi_2+y\eta_2) +(x\xi_1+y\eta_1)^2(x\xi_2+y\eta_2),\\
                &
-\frac{2\lambda_1}{\alpha}(x\xi_1+y\eta_1)^2-2\alpha
(x\xi_1+y\eta_1)(x\xi_2+y\eta_2) -(x\xi_1+y\eta_1)^2(x\xi_2+y\eta_2),
\end{aligned}\right.
\nonumber \\
& 
= \left\{\begin{aligned} 
         & 2\alpha^3[(x+y)^2+\alpha (x^2-y^2)]+\alpha^5[(x+y)(x^2-y^2)-\alpha (x+y)^3],\\
         &-2\alpha^3[(x+y)^2+\alpha (x^2-y^2)]-\alpha^5[(x+y)(x^2-y^2)-\alpha (x+y)^3].
\end{aligned}\right.\nonumber
\end{align}
Because the first eigenvector space
$E_1=\text{span}\{\xi ,\eta\}$ of (\ref{11.84}) with (\ref{11.81}) is
invariant for the equations (\ref{11.79}) with (\ref{11.81}), the
center manifold function $\Phi$ vanishes, i.e.,
$$\Phi (x,y)\equiv 0.$$
Therefore,  we derive  from (\ref{11.112})-(\ref{11.115})  that
\begin{align*}
&\frac{(G(x\xi +y\eta +\Phi ),\xi^*)}{(\xi ,\xi^* ) }=
a_{20}x^2+a_{11}xy+a_{02}y^2+a_{30}x^3+a_{21}x^2y+a_{12}xy^2+a_{03}y^3,\\
&
\frac{(G(x\xi +y\eta +\Phi ),\eta^* )}{(\eta ,\eta^* ) }
=b_{20}x^2+b_{11}xy+b_{02}y^2+b_{30}x^3+b_{21}x^2y+b_{12}xy^2+b_{03}y^3,
\end{align*}
where
\begin{align*}
&a_{20}=\alpha (\alpha +1),  && a_{02}=\alpha (1-\alpha ),\\
& b_{20}=\alpha (\alpha +1), && b_{02}=\alpha (1-\alpha ),\\
&  a_{11}=2\alpha , && b_{11}=2\alpha ,\\
& a_{30}=\frac{1}{2}\alpha^3(1-\alpha ),&& b_{03}=-\frac{1}{2}\alpha^3(1+\alpha ),\\
&  a_{12}=-\frac{1}{2}\alpha^3(1+3\alpha ),&& b_{21}=\frac{1}{2}\alpha^3(1-3\alpha ).
\end{align*}
Then, the parameter $b$ in  Theorem~A.3 in \cite{MW10a} is
\begin{align*}
b
=&\frac{\pi}{2\alpha}(a_{02}b_{02}-a_{20}b_{20})+\frac{\pi}{4\alpha}(a_{11}a_{20}+a_{11}a_{02}
-b_{11}b_{20}-b_{11}b_{02})\\
&+\frac{3\pi}{4}(a_{30}+b_{03})+\frac{\pi}{4}(a_{12}+b_{21})\\
=&-\pi\alpha^2(2+\frac{3}{2}\alpha^2). 
\end{align*} 
Namely,
$b<0$. Hence, by Theorem~A.3 in \cite{MW10a}, the system (\ref{11.83}) bifurcates
from $(v,\lambda )=(0,\lambda_1)$ to a periodic solution on $\lambda
>\lambda_1$, which is an attractor. The proof of the expression
(\ref{11.111}) is  classical.  
Thus the theorem is proved.
\ep

Now, we consider the Dirichlet boundary condition. In this case,
$\rho_1>0$ and $\lambda_1=(\mu_1+\mu_2)\rho_1+\alpha^2+1$. By
(\ref{11.88}) and (\ref{11.89}) it is easy to see that as $\lambda_1<\lambda_0$
we have
$$\mu_2\rho_1(\mu_2\rho_1+\alpha^2)<\alpha^2(\mu_1\rho_1+1).$$
Then we define the following parameter
\begin{align}
b_1=&\frac{2\pi\alpha^2[\int_{\Omega}e^3_1dx]^2}{\sigma^2_0[\int_{\Omega}e^2_1dx]^2}(\mu_1
\rho_1+1)(\mu^2_2\rho^2_1+2\mu_2\rho_1(\mu_1\rho_1+1)-\sigma^2_0)\label{11.116}\\
&-\frac{\pi\alpha^2\int_{\Omega}e^4_1dx}{2\int_{\Omega}e^2_1dx}(2\mu_2\rho_1+3\alpha^2)\nonumber\\
&+\frac{2\pi\alpha^2}{\int_{\Omega}e^2_1dx}[(3\mu_1\rho_1+\mu_2\rho_1+3)A_1+(\mu_1\mu_2
\rho^2_1+\mu_2\rho_1+\delta_0)B_1]\nonumber\\
&-\frac{8\pi\alpha^2\sigma^2_0}{\int_{\Omega}e^2_1dx}[(\mu_1\rho_1+\mu_2\rho_1+1)A_2-(\mu_1
\mu_2\rho^2_1+\mu_2\rho_1-\sigma_0)B_2]\nonumber\\
&-\frac{4\pi\alpha^2}{\int_{\Omega}e^2_1dx}[(\mu_1\mu_2\rho^2_1+\mu_2\rho_1-\sigma^2_0)A_3+
\sigma^2_0(\mu_1\rho_1+\mu_2\rho_1+1)B_3], \nonumber
\end{align}
where
\begin{align*}
&
\sigma_0=[\alpha^2(\mu_1\rho_1+1)-\mu_2\rho_1(\mu_2\rho_1+\alpha^2)]^{{1}/{2}},\\
&
A_i=(2\lambda_1-\mu_2\rho_1-\alpha^2)B_i+\alpha^2C_i \quad \text{ for } i=1,2,3,\\
&
B_1=\sum\limits^{\infty}_{k>1}\frac{\mu_2\rho_k[\int_{\Omega}e^2_1e_kdx]^2}{\text{det}M_k
\int_{\Omega}e^2_kdx},\\
&
B_2=\sum\limits^{\infty}_{k>1}\frac{[\int_{\Omega}e^2_1e_kdx]^2}{\text{det}M_k\text{det}(M^2_k+
4\sigma^2_0)\int_{\Omega}e^2_kdx} [(\mu_2\rho_k+\alpha^2)D_k+\alpha^2Q_k], \\
&
B_3=\sum\limits^{\infty}_{k>1}\frac{[\int_{\Omega}e^2_1e_kdx]^2}{\text{det}(M^2_k+4\sigma^2_0)
\int_{\Omega}e^2_kdx}D_k,\\
&
C_1=-\sum\limits^{\infty}_{k>1}\frac{(\mu_1\rho_1+1)[\int_{\Omega}e^2_1e_kdx]^2}{\text{det}
M_k\int_{\Omega}e^2_kdx},\\
&
C_2=\sum\limits^{\infty}_{k>1}\frac{[\int_{\Omega}e^2_1e_kdx]^2}{\text{det}M_k\text{det}(M^2_k+
4\sigma^2_0)\int_{\Omega}e^2_kdx} [(\mu_1\rho_k+1-\lambda_1)Q_k-\lambda_1D_k], \\
&
C_3=\sum\limits^{\infty}_{k>1}\frac{[\int_{\Omega}e^2_1e_kdx]^2}{\text{det}(M^2_k+4\sigma^2_0)
\int_{\Omega}e^2_kdx}Q_k,\\
&
D_k=(\mu_2\rho_k+\alpha^2)^2+4\sigma^2_0-\alpha^2(\mu_1+\mu_2)\rho_k-\alpha^2(\alpha^2+1),\\
&
Q_k=\lambda_1\alpha^2-\lambda_1(\mu_1+\mu_2)(\rho_k-\rho_1)-(\mu_1\rho_k+1-\lambda_1)^2-4\sigma^2_0.
\end{align*}
Here $M_k$ is the matrix defined by
\begin{equation}
M_k=\left(\begin{array}{cc} -\mu_1\rho_k+\lambda_1-1&\alpha^2\\
-\lambda_1&-\mu_2\rho_k-\alpha^2
\end{array}\right).\label{11.117}
\end{equation}

\bt\la{t11.9}
 Let $b_1$ be the number given by (\ref{11.116}) and
$\lambda_1<\lambda_0$. For the problem (\ref{11.79})-(\ref{11.80}), the following assertions hold true:

\begin{itemize}

\item[(1)] The problem undergoes a dynamic transition at $\lambda =\lambda_1$,
which is the Hopf bifurcation.

\item[(2)] When $b_1<0$, the transition is of Type-I and bifurcates
to a stable periodic solution on $\lambda >\lambda_1$, and when $b_1>0$
the transition is of Type-II and bifurcates to an unstable periodic
solution on $\lambda <\lambda_1$.

\item[(3)] The bifurcated periodic solution
$v^{\lambda}=(v^{\lambda}_1,v^{\lambda}_2)$ can be expressed as
\begin{equation}
\begin{aligned} 
&  
v^{\lambda}_1=2\alpha^2[-\gamma (\lambda)/b_1]^{{1}/{2}}e_1\sin (\sigma_0t+\frac{\pi}{4})
 +o(|\gamma |^{{1}/{2}}),\\
&
v^{\lambda}_2=2(\sigma^2_0+(\mu_2\rho_1+\alpha^2)^2)[-\frac{\gamma
(\lambda )}{b_1}]^{{1}{2}}e_1\cos (\sigma_0t+\theta )+o(|\gamma|^{{1}/{2}}),\\
&
\theta
=\tan^{-1}\frac{\sigma_0+\mu_2\rho_1+\alpha^2}{\sigma_0-\mu_2\rho_1-\alpha^2},
\end{aligned}
\label{(11.118)}
\end{equation}
where $\gamma =(\lambda -\lambda_1)/2.$
\end{itemize}
\et

\bp
 By (\ref{11.86}) the eigenvalues and eigenvectors of
(\ref{11.84}) with (\ref{11.80}) at $\lambda_1= (\mu_1+\mu_2)\rho_1+\alpha^2+1$
are determined by the matrices $M_k$ given by (\ref{11.117}). It is clear
that $M_1$ has a pair of imaginary eigenvalues
$$\beta^{\pm}_1(\lambda_1)=\pm i\sigma_0\ \ \ \ (\sigma_0\ \text{as\
in\ (\ref{11.116})}).$$ 
Let $\tilde{\xi},\tilde{\eta}\in \R^2$ be the
eigenvectors of $M_1$ satisfying
$$M_1\tilde{\xi}=\sigma_0\tilde{\eta},\ \ \ \
M_1\tilde{\eta}=-\sigma_0\tilde{\xi}.$$ 
Then, by (\ref{11.86}) the
eigenvectors of (\ref{11.84}) corresponding to $\beta^{\pm}_1(\lambda_1)$
are given by $\xi =\tilde{\xi}e_1$ and $\eta =\tilde{\eta}e_1$. It
is readily to check that
\begin{align}
&
\xi
=(\xi_1,\xi_2)=(\alpha^2e_1,(\sigma_0-\mu_2\rho_1-\alpha^2)e_1),\label{11.119}
\\
&
\eta
=(\eta_1,\eta_2)=(\alpha^2e_1,-(\sigma_0+\mu_2\rho_1+\alpha^2)e_1).\label{11.120}
\end{align}

We consider the conjugate eigenvectors $\xi^*=\tilde{\xi}^*e_1$ and
$\eta^*=\tilde{\eta}^*e_1$ with
$$M^*_1\tilde{\xi}^*=\sigma_0\tilde{\eta}^*,\ \ \ \
M^*_1\tilde{\eta}^*=-\sigma_0\tilde{\xi}^*,$$ where $M^*_1$ is the
transpose of $M_1$. Direct calculation shows  that
\begin{align}
&
\xi^*=(\xi^*_1,\xi^*_2)=((\sigma_0+\mu_2\rho_1+\alpha^2)e_1,\alpha^2e_1),\label{11.121}
\\
&
\eta^*=(\eta^*_1,\eta^*_2)=((-\sigma_0+\mu_2\rho_1+\alpha^2)e_1,\alpha^2e_1).\label{11.122}
\end{align}
It is easy to see that
\begin{equation}
\left.\begin{array}{l} (\xi ,\eta^*)=(\eta ,\xi^*)=0,\\
(\xi ,\xi^*)=-(\eta ,\eta^*)=2\alpha^2\sigma_0\int_{\Omega}e^2_1dx.
\end{array}\right.\label{11.123}
\end{equation}

Let $u=x\xi +y\eta +\Phi (x,y)\in H$ be a solution of
(\ref{11.79})-(\ref{11.80}) at $\lambda =\lambda_1$, and $\Phi$  be  the center
manifold function. By (\ref{11.123}) the reduced equations of
(\ref{11.79})-(\ref{11.80}) read
\begin{equation}
\begin{aligned}
&  \frac{dx}{dt}=-\sigma_0y+\frac{1}{(\xi,\xi^*)}(G(x\xi +y\eta +\Phi ),\xi^*),\\
& \frac{dy}{dt}=\sigma_0x+\frac{1}{(\eta ,\eta^*)}(G(x\xi +y\eta +\Phi),\eta^*),
\end{aligned} 
\label{11.124}
\end{equation}
where the operator $G$ is given by
\begin{equation}
G(u)=G_2(u)+G_3(u),\label{11.125}
\end{equation}
and $G_k$   $(k=2,3)$ is a $k$-multilinear operator defined by
\begin{equation}
\begin{aligned}
&  G_2(u,v)=2\left(\frac{\lambda_1}{\alpha}u_1v_1+\alpha u_1v_2,
     -\left(\frac{\lambda_1}{\alpha}u_1v_1+\alpha u_1v_2\right)\right),\\
& G_3(u,v,w)=(u_1v_1w_2,-u_1v_1w_2), \\
& G_2(u)=G_2(u, u), \\
& G_3(u)=G_3(u, u, u).
\end{aligned}
\label{11.125-1}
\end{equation}

Based on (\ref{11.119})-(\ref{11.123}) and (\ref{11.125})-(\ref{11.125-1}), 
(\ref{11.124}) are rewritten as
\begin{equation}\label{11.125-2}
\begin{aligned}
\frac{dx}{dt}
=&-\sigma_0y+a_{20}x^2+a_{11}xy+a_{02}y^2+a_{30}x^3+a_{21}x^2y \\
&+a_{12}xy^2+a_{03}y^3+\frac{x}{(\xi ,\xi^*)}(G_2(\xi ,\Phi)+G_2(\Phi ,\xi ),\xi^*) \\
&+\frac{y}{(\xi ,\xi^*)}(G_2(\eta ,\Phi )+G(\Phi ,\eta),\eta^*)+o(3),\\
\frac{dy}{dt}=&\sigma_0x+b_{20}x^2+b_{11}xy+b_{22}y^2+b_{30}x^3+b_{21}x^2y\\
&+b_{12}xy^2+b_{03}y^3+\frac{x}{(\eta ,\eta^*)}(G_2(\xi ,\Phi)+G_2(\Phi ,\xi ),\eta^*)\\
&+\frac{y}{(\eta ,\eta^*)}(G_2(\eta ,\Phi )+G_2(\Phi ,\eta),\eta^*)+o(3),
\end{aligned}
\end{equation}
where
\begin{align*}
a_{20}= & \frac{(G_2(\xi ,\xi ),\xi^*)}{(\xi,\xi^*)}=\frac{\alpha\int_{\Omega}e^3_1dx}{\sigma_0\int_{\Omega}e^2_1dx}(\sigma_0+\mu_2\rho_1)(\mu_1
\rho_1+1+\sigma_0),\\
a_{11}= & \frac{(G_2(\xi ,\eta )+G_2(\eta ,\xi ),\xi^*)}{(\xi,\xi^*)}=\frac{2\alpha\int_{\Omega}e^3_1dx}{\sigma_0\int_{\Omega}e^2_1dx}(\sigma_0+\mu_2\rho_1)
(\mu_1\rho_1+1),\\
a_{02}= & \frac{(G_2(\eta ,\eta ),\xi^*)}{(\xi,\xi^*)}=\frac{\alpha\int_{\Omega}e^3_1dx}{\sigma_0\int_{\Omega}e^2_1dx}(\sigma_0+\mu_2\rho_1)(
\mu_1\rho_1+1-\sigma_0),\\
a_{30}=& \frac{(G_3(\xi ,\xi ,\xi ),\xi^*)}{(\xi
,\xi^*)}=\frac{\alpha^2\int_{\Omega}e^4_1dx}{2\sigma_0\int_{\Omega}e^2_1dx}(\sigma_0+\mu_2\rho_1)
(\sigma_0-\mu_2\rho_1-\alpha^2), \\
a_{21}= & \frac{1}{(\xi ,\xi^*)}(G_3(\xi ,\xi ,\eta )+G_3(\xi ,\eta
,\xi )+(G_3(\eta ,\xi ,\xi ),\xi^*)\\
= & \frac{\alpha^2\int_{\Omega}e^4_1dx}{2\sigma_0\int_{\Omega}e^2_1dx}(\sigma_0+\mu_2\rho_1)(\sigma_0- 3\mu_2\rho_1-3\alpha^2),\\
a_{12}= & \frac{1}{(\xi ,\xi^*)}(G_3(\xi ,\eta ,\eta )+G_3(\eta ,\xi,\eta )+G_3(\eta ,\eta ,\xi ),\xi^*)\\
= & -\frac{\alpha^2\int_{\Omega}e^4_1dx}{2\sigma_0\int_{\Omega}e^2_1dx}(\sigma_0+\mu_2\rho_1)(\sigma_0+ 3\mu_2\rho_1+3\alpha^2),\\
a_{03}= & \frac{(G_3(\eta ,\eta ,\eta ),\xi^*)}{(\xi,\xi^*)}=-\frac{\alpha^2\int_{\Omega}e^4_1dx}{2\sigma_0\int_{\Omega}e^2_1dx}(\sigma_0+\mu_2
\rho_1)(\sigma_0+\mu_2\rho_1+\alpha^2),
\end{align*}
and 
\begin{align*}
b_{20}= & \frac{(G_2(\xi ,\xi ),\eta^*)}{(\eta,\eta^*)}=\frac{\alpha\int_{\Omega}e^3_1dx}{\sigma_0\int_{\Omega}e^2_1dx}(\sigma_0-\mu_2\rho_1)
(\mu_1\rho_1+1+\sigma_0),\\
b_{11}= & \frac{(G_2(\xi ,\eta )+G(\eta ,\xi ),\eta^*)}{(\eta
,\eta^*)}=\frac{2\alpha\int_{\Omega}e^3_1dx}{\sigma_0\int_{\Omega}e^2_1dx}(\sigma_0-\mu_2\rho_1)
(\mu_1\rho_1+1),\\
b_{02}= & \frac{(G_2(\eta ,\eta ),\eta^*)}{(\eta,\eta^*)}=\frac{\alpha\int_{\Omega}e^3_1dx}{\sigma_0\int_{\Omega}e^2_1dx}(\sigma_0-\mu_2
\rho_1)(\mu_1\rho_1+1-\sigma_0),\\
b_{30}= & \frac{(G_3(\xi ,\xi ,\xi ),\eta^*)}{(\eta,\eta^*)}=\frac{\alpha^2\int_{\Omega}e^4_1dx}{2\sigma_0\int_{\Omega}e^2_1dx}(\sigma_0-\mu_2\rho_1)
(\sigma_0-\mu_2\rho_1-\alpha^2),\\
b_{21}= &  \frac{1}{(\eta ,\eta^*)}(G_3(\xi ,\xi ,\eta )+G_3(\xi ,\eta,\xi )+G(\eta ,\xi ,\xi ),\eta^*)\\
= & \frac{\alpha^2\int_{\Omega}e^4_1dx}{2\sigma_0\int_{\Omega}e^2_1dx}(\sigma_0-\mu_2\rho_1)(\sigma_0- 3\mu_2\rho_1-3\alpha^2),\\
b_{12}= & \frac{1}{(\eta ,\eta^*)}(G_3(\eta ,\eta ,\xi )+G_3(\eta,\xi ,\eta )+G_3(\xi ,\eta ,\eta ),\eta^*)\\
=& -\frac{\alpha^2\int_{\Omega}e^4_1dx}{2\sigma_0\int_{\Omega}e^2_1dx}(\sigma_0-\mu_2\rho_1)(\sigma_0 +3\mu_2\rho_1+3\alpha^2),\\
b_{03}  = & \frac{(G_3(\eta ,\eta ,\eta ),\eta^*)}{(\eta,\eta^*)}=-\frac{\alpha^2\int_{\Omega}e^4_1dx}{2\sigma_0\int_{\Omega}e^2_1dx}(\sigma_0-\mu_2
\rho_1)(\sigma_0+\mu_2\rho_1+\alpha^2).
\end{align*}

We are now in a position to derive the center manifold function $\Phi$.
By  (A.10) in \cite{MW09c},  
\begin{equation}
\Phi =\Phi_1+\Phi_2+\Phi_3+o(2),\label{11.126}
\end{equation}
where
\begin{align*}
&
-L_{\lambda_1}\Phi_1=P_2\big[ G_2(\xi ,\xi )x^2+ (G_2(\xi ,\eta)+G_2(\eta ,\xi ))xy
+ G_2(\eta,\eta )y^2 \big], \\
&
-[L^2_{\lambda_1}+4\sigma^2_0]L_{\lambda_1}\Phi_2=2\sigma^2_0P_2
 \big[(G_2(\xi,\xi )-G_2(\eta ,\eta ))(y^2-x^2) - 2(G_2(\xi ,\eta )+G_2(\eta ,\xi))xy\big],\\
& [L^2_{\lambda_1}+4\sigma^2_0 ] \Phi_3=\sigma_0P_2\big[ (G_2(\xi ,\eta)+G_2(\eta ,\xi ))(y^2-x^2) 
+2(G_2(\xi ,\xi )-G_2(\eta ,\eta))xy\big],
\end{align*}
$L_{\lambda_1}=A+B_{\lambda_1}$ is the linear operator defined
by (\ref{11.82}), $P_2:H\rightarrow E_2$ the canonical projection, and 
$E_2=\{u\in H| (u,\xi^*)=0,(u,\eta^*)=0\}$ is the complement 
 of $E_1=\text{span}\{\xi ,\eta\}$ in $H$. Note that the
eigenvectors of $L_{\lambda_1}$ satisfy
\begin{eqnarray*}
&&\phi_k=\tilde{\phi}_ke_1,\ \ \ \ \tilde{\phi}_k\in \R^2,\ \ \ \
k=1,2,\cdots ,\\
&&M_k\tilde{\phi}_k=\beta_k\tilde{\phi}_k\ \ \ \  (M_k\ \text{the\
matrix\ as\ in\ (\ref{11.117})}).
\end{eqnarray*}
Hence, we obtain from (\ref{11.119}),(\ref{11.120}),(\ref{11.125-1}) and (\ref{11.126}) that
\begin{align*}
\Phi_1= 
 & 
2\alpha^3[(\mu_1\rho_1+\sigma_0+1)x^2+2(\mu_1\rho_1+1)xy+(\mu_1\rho_1-\sigma_0+1)y^2]\\
&
\times\sum\limits^{\infty}_{k>1}\frac{\int_{\Omega}e^2_1e_kdx}{\int_{\Omega}e^2_kdx}(-M_k)^{-1}
      \left( \begin{matrix} 1\\ -1\end{matrix}\right)e_k,\\
\Phi_2= & 
8\alpha^3\sigma^2_0[\sigma_0(y^2-x^2)-2(\mu_1\rho_1+1)xy] 
\sum\limits^{\infty}_{k>1}\frac{\int_{\Omega}e^2_1e_kdx}{\int_{\Omega}e^2_kdx}(-M_k)^{-1}
(M^2_k+4\sigma^2_0)^{-1}\left(\begin{matrix} 1\\ -1\end{matrix}\right)e_k,\\
\Phi_3= & 4\alpha^3\sigma_0[(\mu_1\rho_1+1)(y^2-x^2)+2\sigma_0xy]
\sum\limits^{\infty}_{k>1}\frac{\int_{\Omega}e^2_1e_kdx}{\int_{\Omega}e^2_kdx}(M^2_k+
4\sigma^2_0)^{-1}\left(\begin{matrix} 1\\
-1\end{matrix}\right)e_k.
\end{align*}
Direct calculation shows that
\begin{align*}
&(-M_k)^{-1}=\frac{1}{\text{det}M_k}\left[\begin{matrix}
\mu_2\rho_k+\alpha^2&\alpha^2\\
-\lambda_1&\mu_1\rho_k+1-\lambda_1
\end{matrix}\right],\\
&(M^2_k+4\sigma^2_0)^{-1}=\frac{1}{\text{det}(M^2_k+4\sigma^2_0)}\left[\begin{matrix}
(\mu_2\rho_k+\alpha^2)^2+4\sigma^2_0-\lambda_1\alpha^2&\alpha^2(\mu_1+\mu_2)(\rho_k-\rho_1)\\
-\lambda_1(\mu_1+\mu_2)(\rho_k-\rho_1)&(\mu_1\rho_k+1-\lambda_1)^2+4\sigma^2_0-\lambda_1\alpha^2
\end{matrix}\right].
\end{align*}
Thus we have
\begin{eqnarray*}
&&\Phi_1=2\alpha^3[(\mu_1\rho_1+1+\sigma_0)x^2+2(\mu_1\rho_1+1)xy+(\mu_1\rho_1+1-\sigma_0)y^2]
\left(\begin{array}{c} E_1\\
F_1\end{array}\right),\\
&&\Phi_2=8\alpha^3\sigma^2_0[-\sigma_0x^2-2(\mu_1\rho_1+1)xy+\sigma_0y^2]\left(\begin{array}{c}
E_2\\
F_2\end{array}\right),\\
&&\Phi_3=4\alpha^3\sigma_0[-(\mu_1\rho_1+1)x^2+2\sigma_0xy+(\mu_1\rho_1+1)y^2]\left(\begin{array}{c}
E_3\\
F_3\end{array}\right),
\end{eqnarray*}
where
\begin{eqnarray*}
&&E_1=\sum\limits^{\infty}_{k>1}\frac{\int_{\Omega}e^2_1e_kdx}{\text{det}M_k\int_{\Omega}e^2_kdx}
\mu_2\rho_ke_k,\\
&&F_1=-\sum\limits^{\infty}_{k>1}\frac{\int_{\Omega}e^2_1e_kdx}{\text{det}M_k\int_{\Omega}e^2_kdx}
(\mu_1\rho_k+1)e_k,\\
&&E_2=\sum\limits^{\infty}_{k>1}\frac{\int_{\Omega}e^2_1e_kdx}{\text{det}M_k\text{det}(M^2_k+4
\sigma^2_0)\int_{\Omega}e^2_kdx}((\mu_2\rho_k+\alpha^2)D_k+\alpha^2Q_k)e_k,\\
&&F_2=\sum\limits^{\infty}_{k>1}\frac{\int_{\Omega}e^2_1e_kdx}{\text{det}M_k\text{det}(M^2_k+4
\sigma^2_0)\int_{\Omega}e^2_kdx}((\mu_1\rho_k+1-\lambda_1)Q_k-\lambda_1D_k)e_k,\\
&&E_3=\sum\limits^{\infty}_{k>1}\frac{\int_{\Omega}e^2_1e_kdx}{\text{det}(M^2_k+4\sigma^2_0)\int_{\Omega}
e^2_kdx}D_ke_k,\\
&&F_3=\sum\limits^{\infty}_{k>1}\frac{\int_{\Omega}e^2_1e_kdx}{\text{det}(M^2_k+4\sigma^2_0)\int_{\Omega}
e^2_kdx}Q_ke_k.
\end{eqnarray*}
Inserting $\Phi =\Phi_1+\Phi_2+\Phi_3+o(2)$ into (\ref{11.125-2}) we
derive  that
\begin{equation}
\begin{aligned}
&  \frac{dx}{dt}=-\sigma_0y+\sum\limits_{2\leq
i+j\leq 3}a_{ij}x^iy^j+\sum\limits_{k+r=3}\tilde{a}_{kr}x^ky^r+o(3), \\
 & \frac{dy}{dt}=\sigma_0x+\sum\limits_{2\leq i+j\leq
3}b_{ij}x^iy^j+\sum\limits_{k+r=3}\tilde{b}_{kr}x^ky^r+o(3),
\end{aligned}
\label{11.127}
\end{equation}
where $a_{ij}$ and $b_{ij}$  $(0\leq i,j\leq 3)$ are as in (\ref{11.125-2}), and
\begin{eqnarray*}
&&\tilde{a}_{30}=\frac{2\alpha^2(\sigma_0+\mu_2\rho_1)}{\sigma_0\int_{\Omega}e^2_1dx}[(\mu_1
\rho_1+1+\sigma_0)I_1-4\sigma^3_0I_2-2\sigma_0(\mu_1\rho_1+1)I_3],\\
&&\tilde{a}_{12}=\frac{2\alpha^2(\sigma_0+\mu_2\rho_1)}{\sigma_0\int_{\Omega}e^2_1dx}[(\mu_1
\rho_1+1-\sigma_0)I_1+4\sigma^3_0I_2+2\sigma_0(\mu_1\rho_1+1)I_3\\
&&\ \ \ \ \ \
+2(\mu_1\rho_1+1)J_1-8\sigma^2_0(\mu_1\rho_1+1)J_2+4\sigma^2_0J_3],\\
&&\tilde{b}_{03}=\frac{2\alpha^2(\sigma_0-\mu_2\rho_1)}{\sigma_0\int_{\Omega}e^2_1dx}[(\mu_1
\rho_1+1-\sigma_0)J_1+4\sigma^3_0J_2+2\sigma_0(\mu_1\rho_1+1)J_3],\\
&&\tilde{b}_{21}=\frac{2\alpha^2(\sigma_0-\mu_2\rho_1)}{\sigma_0\int_{\Omega}e^2_1dx}[2(\mu_1
\rho_1+1)I_1-8\sigma^2_0(\mu_1\rho_1+1)I_2+4\sigma^2_0I_3\\
&&\ \ \ \ \ \
+(\mu_1\rho_1+1+\sigma_0)J_1-4\sigma^3_0J_2-2\sigma_0(\mu_1\rho_1+1)J_3],
\end{eqnarray*}
where
\begin{eqnarray*}
&&I_i=A_i+\sigma_0-\int_{\Omega}E_ie^2_1dx,\\
&&J_i=A_i-\sigma_0\int_{\Omega}E_ie^2_1dx,\\
&&A_i=(2\lambda_1-\mu_2\rho_1-\alpha^2)\int_{\Omega}E_ie^2_1dx+\alpha^2\int_{\Omega}F_ie^2_1dx,
\ \ \ \ i=1,2,3.
\end{eqnarray*}
Then, the number
\begin{align*}
b_1=&\frac{\pi}{2\sigma_0}(a_{02}b_{02}-a_{20}b_{20})+\frac{\pi}{4\sigma_0}(a_{11}a_{20}+a_{11}a_{02}-
b_{11}b_{20}-b_{11}b_{02})\\
&+\frac{3\pi}{4}(a_{30}+b_{30}+\tilde{a}_{30}+\tilde{b}_{30})+\frac{\pi}{4}(a_{12}+b_{21}+
\tilde{a}_{12}+\tilde{b}_{21})
\end{align*}
is the same as in (\ref{11.116}). Thus Assertions (1)-(2) of this theorem
follows from Theorem A.6 in \cite{MW09b}. 

It is known that the bifurcated periodic solution near $\lambda
=\lambda_1$ takes the form
\begin{equation}
v^{\lambda}=x(t)\xi +y(t)\eta +o(|x|+|y|),\label{11.128}
\end{equation}
where $\xi ,\eta$ are as in (\ref{11.119}) and (\ref{11.120}), and $(x(t),y(t))$
is the solution of the following equation
\begin{eqnarray*}
&&\frac{dx}{dt}=\gamma (\lambda )x-\sigma_0(\lambda
)y+\frac{1}{(\xi_{\lambda},\xi^*_{\lambda})}(G(x\xi_{\lambda}+y\eta_{\lambda}+\Phi_{\lambda}),\xi^*_{\lambda}),\\
&&\frac{dy}{dt}=\sigma_0(\lambda )x+\gamma (\lambda
)y+\frac{1}{(\eta_{\lambda},\eta^*_{\lambda})}(G(x\xi_{\lambda}+y\eta_{\lambda}+\Phi_{\lambda}),
\eta^*_{\lambda}),
\end{eqnarray*}
where $\xi_{\lambda},\eta_{\lambda}$ are eigenvectors of
$L_{\lambda}$ corresponding to the first complex eigenvalues
$\beta^{\pm}_1=\gamma\pm i\sigma_0$, and
$\xi^*_{\lambda},\eta^*_{\lambda}$ the conjugate eigenvectors. This
solution $(x(t),y(t))$ near $\lambda_1$ is of the form
\begin{equation}
\left.\begin{array}{l} x(t)=\left[-\frac{\gamma (\lambda
)}{b_1}\right]^{{1}/{2}}\cos\sigma_0t+o(|\gamma (\lambda )|^{1/2}),\\
y(t)=\left[-\frac{\gamma (\lambda
)}{b_1}\right]^{{1}/{2}}\sin\sigma_0t+o(|\gamma (\lambda )|^{1/2}),
\end{array}\right.\label{11.129}
\end{equation}
where $b_1$ is as in (\ref{11.116}). Therefore, Assertion (3) follows from
(\ref{11.128}) and (\ref{11.129}). The proof is complete.
\ep

\section{One-dimensional case}
When the containers $\Omega$ are taken as rectangles, the 
criteria  in Theorems \ref{t11.7} and \ref{t11.9} can be simplified. For simplicity,
 we consider  here only the one-dimensional
case: 
\begin{equation}
\Omega =(0,L)\subseteq \R^1.\label{11.130}
\end{equation}

The eigenvalues $\rho_k$  and  corresponding eigenvectors of (\ref{11.85}) are given by 
\begin{align}
&
\rho_k=\left\{
\begin{aligned} 
   &  k^2\pi^2/L^2    &&\text{for\ b.c.\ (\ref{11.80})},\\
   &  (k-1)^2\pi^2/L^2 &&\text{for\ b.c.\ (\ref{11.81})},
\end{aligned}\right.\ \ \ \ (k=1,2,\cdots ),\label{11.131}
\\
&
e_k=\left\{\begin{aligned} 
  & \sin\frac{k\pi x}{L} &&\text{for\ b.c.\ (\ref{11.80}),}\\
  & \cos\frac{(k-1)\pi x}{L} &&\text{for\ b.c.\ (\ref{11.81}).}
\end{aligned}
\right.\label{11.132}
\end{align}
Thus the  two critical numbers $\lambda_0$ and $\lambda_1$ in
(\ref{11.88}) and (\ref{11.89}) are given by 
\begin{align}
& 
\lambda_0=\text{min}_{k^2}\left[\frac{\mu_1k^2\pi^2}{L^2}+\frac{\alpha^2L^2}{\mu_2k^2\pi^2}+
\frac{\mu_1\alpha^2}{\mu_2}+1\right],\label{11.133}
\\
&
\lambda_1=\left\{\begin{aligned}
&  \frac{\pi^2}{L^2}(\mu_1+\mu_2)+\alpha^2+1 &&\text{for\ b.c.\
(\ref{11.80})},\\
& \alpha^2+1 &&\text{for\ b.c.\ (\ref{11.81}).}
\end{aligned}\right.\label{11.134}
\end{align}
It is known that the criterion $b_1$ in Theorem \ref{t11.7} is valid only
for the free boundary condition, which can be expressed explicitly
by (\ref{11.110}). Likewise, for the number defined by (\ref{11.116}) we have
the following explicit expression
\begin{equation}
b_1=\frac{2\pi\alpha^2}{L}b_0,\label{11.135}
\end{equation}
with
\begin{eqnarray}
b_0&=&\frac{4^3\times
L}{9\sigma^2_0\pi^2}(\mu_1\rho_1+1)(\mu^2_2\rho^2_1+2\mu_1\mu_2\rho^2_1+2\mu_2\rho_1-\sigma^2_0)\label{11.136}\\
&&-\frac{3L}{16}(2\mu_2\rho_1+3\alpha^2)+2(3\mu_1\rho_1+\mu_2\rho_1+3)A_1\nonumber\\
&&-8\sigma^2_0(\mu_1\rho_1+\mu_2\rho_1+1)A_2-4(\mu_1\mu_2\rho^2_1+\mu_2\rho_1-\sigma^2_0)A_3\nonumber\\
&&+2(\mu_1\mu_2\rho^2_1+\mu_2\rho_1+\sigma_0)B_1+8\sigma^2_0(\mu_1\mu_2\rho^2_1+\mu_2\rho_1-\sigma_0)
B_2\nonumber\\
&&-4\sigma^2_0(\mu_1\rho_1+\mu_2\rho_1+1)B_3,\nonumber
\end{eqnarray}
where
$A_i=(2\mu_1\rho_1+\mu_2\rho_1+\alpha^2+2)B_i+\alpha^2C_i$   $(i=1,2,3)$,
and
\begin{eqnarray*}
&&B_1=32L\sum\limits^{\infty}_{k=1}\frac{\mu_2}{\text{det}M_{2k+1}[(2k+1)^2-4]^2L^2},\\
&&B_2=33L\sum\limits^{\infty}_{k=1}\frac{(\mu_2\rho_{2k+1}+\alpha^2)D_{2k+1}+\alpha^2Q_{2k+1}}{\text{det}
M_{2k+1}\text{det}(M^2_{2k+1}+4\sigma^2_0)\pi^2(2k+1)^2[(2k+1)^2-4]^2},\\
&&B_3=32L\sum\limits^{\infty}_{k=1}\frac{D_{2k+1}}{\text{det}(M^2_{2k+1}+4\sigma^2_0)\pi^2(2k+1)^2
[(2k+1)^2-4]^2},\\
&&C_1=-32L\sum\limits^{\infty}_{k=1}\frac{\mu_1\rho_1+1}{\text{det}M_{2k+1}\pi^2(2k+1)^2[(2k+1)^2-4]^2},\\
&&C_2=-32L\sum\limits^{\infty}_{k=1}\frac{(\mu_1\rho_1+\mu_2\rho_1+\alpha^2+1,D_{2k+1}-(\mu_1
\rho_k-\mu_1\rho_1-\mu_2\rho_1-\alpha^2)Q_{2k+1}}{\text{det}M_{2k+1}\text{det}(M^2_{2k+1}+4\sigma^2_0)
\pi^2(2k+1)^2[(2k+1)^2-4]^2]},\\
&&C_3=32L\sum\limits^{\infty}_{k=1}\frac{Q_{2k+1}}{\text{det}(M^2_{2k+1}+4\sigma^2_0)\pi^2(2k+1)^2
[(2k+1)^2-4]^2}.
\end{eqnarray*}

Let $\lambda_0$ in (\ref{11.133}) achieves its minimum at the integer
$k^2_0$, i.e., 
\begin{equation}
\lambda_0=\frac{\mu_1k^2_0\pi^2}{L^2}+\frac{\alpha^2L^2}{\mu_2k^2_0\pi^2}+\frac{\mu_1}{\mu_2}\alpha^2+1.
\label{11.137}
\end{equation}
Then  $k_0\geq 1$ satisfies
\begin{equation}
k_0(k_0-1)\leq\frac{\alpha L^2}{\pi^2\sqrt{\mu_1\mu_2}}\leq
k_0(k_0+1).\label{11.138}
\end{equation}
To see this, note  that the function $$\lambda
(x)=\frac{\mu_1\pi^2x}{L^2}+\frac{\alpha^2L^2}{\mu_2\pi^2x}+\frac{\mu_1}{\mu_2}\alpha^2+1$$
has its minimum at $x_0=\alpha L^2/\pi^2\sqrt{\mu_1\mu_2}$, and
$$\frac{d\lambda}{dx}
\left\{\begin{aligned}
& <0 &&\text{ if }\ x<x_0,\\
& >0 &&\text{ if }\ x>x_0.
\end{aligned}\right.
$$
It follows that  either $k_0=m$  or $k_0= m+1,$
such that $m=\sqrt{x_0}-\varepsilon$ for some $0<\varepsilon <1$; 
namely
$$m^2\leq x_0\leq (m+1)^2.$$
It follows that
\begin{equation}
k_0=\left\{\begin{aligned} 
& m &&\text{ if }\ \lambda (m^2)<\lambda ((m+1)^2),\\
& m\ \text{and}\ m+1   &&\text{ if }\ \lambda (m^2)=\lambda ((m+1)^2),\\
& m+1 && \text{ if }\ \lambda (m^2)>\lambda ((m+1)^2).
\end{aligned}\right.\label{11.139}
\end{equation}
We infer from (\ref{11.139}) that
\begin{eqnarray*}
&&\frac{\alpha L^2}{\pi^2\sqrt{\mu_1\mu_2}}\leq m(m+1)\Rightarrow
k_0=m,\\
&&\frac{\alpha L^2}{\pi^2\sqrt{\mu_1\mu_2}}\geq m(m+1)\Rightarrow
k_0=m+1,
\end{eqnarray*}
which yield the inequalities (\ref{11.138}).

In the following, we compare $\lambda_0$ with $\lambda_1$ in terms of the  
parameters $\mu_1,\mu_2,\alpha$, and $L$. We proceed in two cases.

\medskip

{\sc The case where $\mu_1\geq\mu_2$:}  Then, from (\ref{11.88})-(\ref{11.89}), we see  that
\begin{equation}
\lambda_1<\lambda_0 \ \ \ \ \forall L,\alpha >0\ \text{for\ b.c.\
(\ref{11.81}).}\label{11.140}
\end{equation}
For b.c. (\ref{11.80}), we can prove that
\begin{equation}
\left.\begin{aligned} 
&  \lambda_0> \lambda_1  &&\text{ if  }\  0<L<L_c,\\
& \lambda_1<\lambda_0,&\text{ if }\ L>L_c,
\end{aligned}\right.\label{11.141}
\end{equation}
where
\begin{equation}
L^2_c=\frac{\pi^2}{2\alpha}[\sqrt{(\mu_1-\mu_2)^2\alpha^2+4\mu^2_2}-\alpha
(\mu_1-\mu_2)].\label{11.142}
\end{equation}
In fact, from $\lambda_0=\lambda_1$, we derive the critical scale
$L_c$ as
$$L^2_c=\frac{k^2_0\pi^2}{2\alpha}\left[
\sqrt{ (\mu_1-\mu_2)^2\alpha^2+\frac{4\mu_2}{k^2_0}(\mu_1+\mu_2
-k^2_0\mu_1)} -\alpha (\mu_1-\mu_2) \right] ,
$$ 
provided that
$$\mu_1+\mu_2>k^2_0\mu_1.$$
Since $\mu_1\geq\mu_2$, it implies that $k_0=1$. Hence $L_c$ is as
in (\ref{11.142}), and (\ref{11.141}) holds true.

We remark that by (\ref{11.138}), $L^2_c$ in (\ref{11.142}) has to satisfy
the inequality
$$L^2_c\leq 2\pi^2\sqrt{\mu_1\mu_2}/\alpha, $$
which holds true for $\mu_1\geq\mu_2$.

{\sc The case where $\mu_1<\mu_2$:}
 For b.c. (\ref{11.80}),  we deduce  from
$\lambda_0=\lambda_1$  the following  two critical scales:
\begin{equation}
L^2_{c_{1,2}}=\frac{k^2_0\pi^2}{2\alpha} \left[
(\mu_2-\mu_1)\alpha\mp\sqrt{(\mu_2-\mu_1)^2\alpha^2-
\frac{4\mu_2}{k^2_0}(k^2_0\mu_1-\mu_1-\mu_2)} \right],\label{(11.142-1)}
\end{equation}
with $L^2_{c_1}<L^2_{c_2}$. 

It is easy to see that  for the boundary condition (\ref{11.80}), 
\begin{align}
&
\lambda_1<\lambda_0 
 && \text{ if } 
    \left\{\begin{aligned}
       &  k^2_0\alpha^2(\mu_2-\mu_1)^2<4\mu_2(k^2_0\mu_1-\mu_1-\mu_2),\\
       & \text{ or } \\
       &   k^2_0\alpha^2(\mu_2-\mu_1)^2>4\mu_2(k^2_0\mu_1-\mu_1-\mu_2) \text{ and }  \\ 
       & \qquad    0<L^2<L^2_{c_1}  \text{ or }\ L^2_{c_2}<L^2, 
     \end{aligned}\right.\label{11.143}
\\
& 
\lambda_0<\lambda_1
  &&  \text{  if } 
     \left\{\begin{aligned} 
        & L^2_{c_1}<L^2<L^2_{c_2}  &&  \text{ for } 0<4\mu_2(k^2_0\mu_1-\mu_1 \mu_2)
             <k^2_0\alpha^2(\mu_2-\mu_1), \\
        & \text{or}\ 0<L^2<L^2_{c_2}  && \text{ for }\ k^2_0\mu_1\leq\mu_1+\mu_2.
       \end{aligned}\right.\label{11.144}
\end{align}
For the boundary condition (\ref{11.81}),  we obtain two critical scales as
\begin{equation}
l^2_{c_{1,2}}=\frac{k^2_0\pi^2}{2\alpha} \left[(\mu_2-\mu_1)\alpha\mp\sqrt{(\mu_2-\mu_1)^2\alpha^2-
4\mu_1\mu_2}\right],\label{11.144-1}
\end{equation}
such that
\begin{align}
& \lambda_1<\lambda_0 \text{ if}\
\left\{\begin{array}{l} (\mu_2-\mu_1)^2\alpha^2<4\mu_1\mu_2,\\
\text{or}\ 0<L^2<l^2_{c_1},\ \text{or}\ l^2_{c_2}<L^2,\\
\text{for}\ (\mu_2-\mu_1)^2\alpha^2>4\mu_1\mu_2,
\end{array}\right.\label{11.145}
\\
&
\lambda_0<\lambda_1\ \text{for\ b.c.\ (\ref{11.81})\ if}\
l^2_{c_1}<L^2<l^2_{c_2}\ \text{for}\
(\mu_2-\mu_1)^2\alpha^2>4\mu_1\mu_2.\label{11.146}
\end{align}

\section{Physical Remarks}

We now  discuss the phase transition of Brusselator by using Theorem
\ref{t11.6}-\ref{t11.9}  for  the one-dimensional case (\ref{11.130}).

\medskip

{\sc Dirichlet Boundary Condition.}
When $\mu_1\geq\mu_2$, by (\ref{11.141}),  the system (\ref{11.79})-(\ref{11.80}) has a
transition to steady states provided $0<L<L_c$, and to periodic
solutions provided $L_c<L$.

\bcon\la{pc11.3}
Let $\mu_1\geq\mu_2$. Then for the
system (\ref{11.79})-(\ref{11.80}),  we have the following conclusions:

\begin{itemize}

\item[(1)] When $0<L<L_c$, the transition of (\ref{11.79})-(\ref{11.80}) at
$\lambda =\lambda_0$ is of Type-III, and there is a saddle-node
bifurcation at some $0<\lambda^*<\lambda_0$. In other wards, the
basic state $u_0=(\alpha ,\lambda /\alpha )$ is stable for
$0<\lambda <\lambda^*$, is metastable for $\lambda^*<\lambda
<\lambda_0$, and is unstable  for $\lambda^*<\lambda$. Moreover, if
$\lambda^*<\lambda$,  there are at least two metastable equilibrium
states.

\item[(2)]  When $L_c<L$, this system undergoes a dynamic transition  at $\lambda
=\lambda_1$ to periodic solutions. In particular, there exists an
$L_0>L_c$ such that if  $L_c<L<L_0$,  the transition is of Type-II, and
there is a singular separation of periodic solutions at some
$\tilde{\lambda}<\lambda_1$. If $L_0<L$,  the transition is of
Type-I,  and the system bifurcates from $u_0$ to a stable periodic solution
on $\lambda >\lambda_1$, which is expressed as
$u_{\lambda}=(u^{\lambda}_1,u^{\lambda}_2)$ with
\begin{eqnarray*}
&&u^{\lambda}_1=\alpha +2\alpha^2\sqrt{\frac{\lambda
-\lambda_1}{|b_1|}}\sin\frac{\pi x}{L}\sin
(\sigma_0t+\frac{\pi}{4})+o(|\lambda -\lambda_1|^{{1}/{2}}),\\
&&u^{\lambda}_2=\frac{\lambda}{\alpha}+2(\sigma^2_0+(\alpha^2+\frac{\mu\pi^2}{L^2})^2)\sqrt{\frac{\lambda
-\lambda_1}{|b_1|}}\sin\frac{\pi x}{L}\cos (\sigma_0t+\theta
)+o(|\lambda -\lambda_1|^{{1}/{2}}).
\end{eqnarray*}
This periodic solution provides a spatial-temporal oscillation of the Brusselator.
\end{itemize}
\econ

The first conclusion is due to Theorem \ref{t11.6}, the existence of global
attractors, and the fact that $u_0=(0,0)$ is a unique steady state
solution of (\ref{11.79})-(\ref{11.80}) at $\lambda =0$.

The second conclusion is based on Theorem \ref{t11.9} and the following
analysis on the criterion $b_0$ given by (\ref{11.136}). We know that
$$\lambda_1-\lambda_0\rightarrow 0\ \ \ \ \text{as}\ \ \ \
L\rightarrow L_c,$$ which implies
$$\sigma_0\rightarrow 0\ \ \ \ \text{as}\ \ \ \ L\rightarrow L_c.$$
It follows from  (\ref{11.136}) that $b_0\rightarrow +\infty$ for
$L\rightarrow L_c+0$. Therefore
\begin{equation}
b_0>0\qquad  \forall L_c<L<L_0,\label{11.147}
\end{equation}
for some $L_0>L_c$. On the other hand, $\rho_k\rightarrow
0$  $(L\rightarrow\infty )$. Hence, when $L\rightarrow\infty$, 
\begin{align}
b_0\rightarrow &
 -\frac{64L}{9\pi^2}-\frac{9\alpha^2L}{16}+2(3\alpha^2+6+\sigma_0)B_1-8\sigma^2_0(
\alpha^2+\sigma_0+2)B_2\label{11.148}\\
&+4\alpha^2(\alpha^2+1)B_3+6\alpha^2C_1-8\sigma^2_0\alpha^2C_2+4\alpha^4C_3.\nonumber
\end{align}
Note that 
\begin{align*}
& \sigma_0\rightarrow\alpha, &&D_k\rightarrow 3\alpha^2,&&Q_k\rightarrow -3\alpha^2,\\
& \text{det}M_k\rightarrow\alpha^2,  &&\text{det}(M^2_k+4\sigma^2_0)\rightarrow
9\alpha^4,&&B_1\rightarrow 0,\\
& B_2\rightarrow 0,  &&B_3\rightarrow\frac{32L}{3\alpha^2}E, &&C_1\rightarrow
-\frac{32L}{\alpha^2}E,\\
&  C_2\rightarrow -\frac{32L}{3\alpha^4}E,&&C_3\rightarrow-\frac{32L}{3\alpha^2}E,
\end{align*}
for $L\rightarrow\infty$, where
$$E=\sum\limits^{\infty}_{k=1}\frac{1}{\pi^2(2k+1)^2[(2k+1)^2-4]^2}.$$
Thus, in view of (\ref{11.148}),
$$b_0\rightarrow -L[\frac{64}{9\pi^2}+\frac{9\alpha^2}{16}+64E]<0 \
\ \ \ \text{as}\ L\rightarrow\infty .
$$ 
Hence 
\begin{equation}
b_0<0 \ \ \ \ \forall L_1<L,\label{11.149}
\end{equation}
for some $L_1 \ge L_0.$

From the physical point of  view, it is reasonable to consider the case where  $b_0$
changes its sign only once in  $(L_c, \infty)$. Hence, physically,  we have $L_0=L_1$. 
Thus, we derive from (\ref{11.147}) and (\ref{11.149}) the second conclusion.

Now, we consider the case where $\mu_1<\mu_2$ by the following two
examples. We take
\begin{equation}
\mu_1=2\times 10^{-3},\ \ \ \ \mu_2=4\times 10^{-3}.\label{11.150}
\end{equation}

\begin{ex}
\la{e11.1}
{\rm
 Let (\ref{11.150}) hold true,  and $\alpha =2,L=4$. Then we obtain 
from (\ref{11.138})  that $k_0=34$, and
$$k^2_0\alpha^2(\mu_2-\mu_1)^2<4\mu_2(k^2_0\mu_1-\mu_1-\mu_2).$$
In view of (\ref{11.143}),
$$\lambda_1<\lambda_0\ \ \ \text{for\ b.c.\ (\ref{11.80}).}$$
The number $b_0$ in (\ref{11.136}) is given by
$$b_0\cong
-\left(\frac{64}{9\pi^2}+\frac{9}{4}\right)L+40B_1-256B_2+80B_3+24C_1-128C_2+64C_3,$$
and
\begin{align*}
&B_1\cong\frac{2}{25}\times 10^{-3}L, &&B_2\cong
-\frac{11}{36\times 25}\times 10^{-3}L,\\
&B_3\cong\frac{8}{27\times 25\pi^2}L, &&C_1\cong -\frac{8}{9\times
25\pi^2}L,\\
&C_2\cong -\frac{2}{27\times 25\pi^2}L, &&C_3\cong -\frac{8}{27\times
25\pi^2}L.
\end{align*}
It is easy to see that $b_0<0$. Then,  by Theorem \ref{t11.9}, the phase
transition of (\ref{11.79})-(\ref{11.80}) is of Type-I, and this system
undergoes a spatial-temporal oscillation on $\lambda
>\lambda_1$.
}
\end{ex}

\begin{ex}\la{e11.2} 
{\rm
Assume (\ref{11.150}) and $\alpha =3,L=4$. Then $k_0=41$, and
$$k^2_0\alpha^2(\mu_2-\mu_1)^2>4\mu_2(k^2_0\mu_1-\mu_1-\mu_2)>0.$$
Thus, we can obtain two critical scales $L^2_{c_1}$ and $L^2_{c_2}$
in (\ref{(11.142-1)}) as follows:
$$L^2_{c_1}=11.07,\ \ \ \ L^2_{c_2}=22.14.$$
Hence 
$$L^2_{c_1}<L^2=16<L^2_{c_2},$$
which leads (by (\ref{11.144})) to
\begin{equation}
\lambda_0<\lambda_1\ \ \ \ \text{for\ b.c.\ (\ref{11.80})}.\label{11.151}
\end{equation}
It is clear that
$$\int_{\Omega}e^3_{k_0}dx=\int^L_0\sin^3\frac{k_0\pi x}{L}dx\neq
0.$$ 
By (\ref{11.151}) and Theorem \ref{t11.6}, the system (\ref{11.79})  and (\ref{11.80}) has a
Type-III transition at $\lambda =\lambda_0$, and there is a
saddle-node bifurcation at some $\lambda^*(0<\lambda^*<\lambda_0)$.
}
\end{ex}

\medskip

{\sc Neumann Boundary Condition.}
Consider the case where $\mu_1\geq\mu_2$. Based on (\ref{11.143}) and
Theorem \ref{t11.8}, we have the following physical conclusion.

\bcon\la{pc11.4} 
Let $\mu_1\geq\mu_2$. Then the
system (\ref{11.79}) with (\ref{11.81}) has a Type-I transition to periodic
solutions at $\lambda =\lambda_1$, i.e., a spatial-temporal oscillation
occurs in the Brusselator for $\lambda >\lambda_1$.
\econ

For the case where $\mu_1<\mu_2$, we have  the following
example.

\begin{ex}
\la{e11.3}
{\rm
Under the same  conditions  as in Example~\ref{e11.2},  
$k_0=41$ and the two critical scales $l^2_{c_1}$ and
$l^2_{c_2}$ in (\ref{11.144-1}) are given by
$$l^2_{c_1}=11.06,\ \ \ \ l^2_{c_2}=22.12.$$
Hence, $l^2_{c_1}<l^2=16<l^2_{c_2}$, which implies, by (\ref{11.146}),
that
\begin{equation}
\lambda_0<\lambda_1\ \ \ \ \text{for\ b.c.\ (\ref{11.81}).}\label{11.152}
\end{equation}
On the other hand, by (\ref{11.131}) and (\ref{11.132}), we have
$$\rho_{k_0+1}=\frac{k^2_0\pi^2}{L^2}\cong 10^3,\ \ \ \
\int_{\Omega}e^3_{k_0+1}dx=\int^L_0\cos^3\frac{k_0\pi x}{L}dx=0.
$$
Thus, it is easy to check that the number $b_1$ in (\ref{11.101}), which
is also given by (\ref{11.110}), is negative, i.e., 
\begin{equation}
b_1<0\ \ \ \ \text{in\ (\ref{11.101})}.\label{11.153}
\end{equation}
By (\ref{11.152})-(\ref{11.153}) and Theorem \ref{t11.7}, the system (\ref{11.79}) with
(\ref{11.81}) bifurcates on $\lambda >\lambda_0$ to two stable steady
states $v^{\lambda}_{\pm}$ as given by (\ref{11.103}). It shows that the
Brusselator undergoes a transition at $\lambda_0=9.8$.
}
\end{ex}

\bibliographystyle{siam}

\end{document}